\documentstyle[aps,prb,preprint]{revtex}
\tightenlines
\title{Correlation between the residual resistance ratio and 
magnetoresistance in MgB$_{2}$}
\author{X. H. Chen, Y. S. Wang, Y. Y. Xue, R. L. Meng, Y. Q. Wang and C. 
W. Chu$^1$}
\address{Department of Physics and Texas Center for Superconductivity, 
University of Houston,\\Houston, Texas 77204-5002, USA\\$^1$also at 
Lawrence Berkeley National Laboratory, 1 Cyclotron Road, Berkeley, 
California 94720; and Hong Kong University of Science and Technology, Hong 
Kong\\submitted July 6, 2001}

\date{\today}

\begin{document}

\maketitle

\begin{abstract}
The resistivity and magnetoresistance in the normal state for
bulk and thin-film MgB$_{2}$ with different nominal compositions
have been studied systematically. These samples show different
temperature dependences of normal state resistivity and residual
resistance ratios although their superconducting transition
temperatures are nearly the same, except for the thin-film sample.
The correlation between the residual resistance ratio (RRR) and
the power law dependence of the low temperature resistivity,
$\rho$ {\it vs.} $T^{c}$, indicates that the electron-phonon interaction
is important. It is found that the magnetoresistance (MR) in the normal 
state scales well with RRR,
$a_0$(MR) $\propto$ (RRR)$^{2.2\pm 0.1}$ at 50~K. This 
accounts for the
large difference in magnetoresistance reported by various groups,
due to different defect scatterings in the samples.
\end{abstract}


\section*{Introduction}

The discovery of superconductivity in MgB$_{2}$ at temperatures as
high as 40~K has attracted considerable interest\cite{nagamatsu}
because of its being a simple inter-metallic binary compound with
negligible grain boundary effect and small anisotropy and thus
its suitability for device applications. The appearance of
superconductivity in MgB$_{2}$ with such a high transition
temperature immediately raises the possibility of even higher
superconducting temperatures in conventional metallic binary
materials. The underlying mechanism of superconductivity in this
system is still an open question. At least two competing
models\cite{kortus,hirsch} have been proposed to account for the
superconducting properties in MgB$_{2}$ and the high $T_c$ of 40~K.
While both models attribute the superconductivity to the
boron-sublattice conduction bands, the pairing mechanisms proposed
differ significantly. Kortus \textit{et al.}~proposed a well-established
phonon-mediated BCS theory, in which the high $T_c$ value is believed to
be due to the high phonon frequencies of the light boron atoms and the
strong electron-phonon interactions. This mechanism is supported
by a number of experiments, such as: isotope effect,\cite{budko}
a strong negative pressure coefficient of
$T_c$,\cite{lorenz,iwasa} quasi-particle
tunneling,\cite{schmidt,sharoni} specific heat,\cite{kremer}
photoemission spectroscopy,\cite{taka} and inelastic neutron
scattering.\cite{yidirim} Alternatively, Hirsch proposed a
``universal'' mechanism in which the superconductivity in MgB$_{2}$ is driven
by the pairing of dressed holes. In fact, an indication of hole-type
conduction in the normal phase was found in the positive
thermoelectric power.\cite{lorenz} The hole character of carriers
was confirmed by Hall measurements.\cite{kang}

Soon after the discovery of superconductivity in MgB$_{2}$, the
resistivity and mangetoresistance in the normal state were
studied.\cite{finnemore,jung,takano,budko1} However, the results
reported differ from group to group. The reported resistivity at
room temperature ranges from 9.6 to 100 $\mu\Omega$cm, and that
at 40~K from 0.38 to 21 $\mu\Omega$cm.\cite{finnemore,jung} Ames
Laboratory\cite{finnemore,budko1} reported that the resistivity in
the normal state followed $T^3$ behavior and observed a large
magnetoresistance, while Jung \textit{et al.}\cite{jung} and
Takano \textit{et al.}\cite{takano} claimed that the resistivity in the
normal state followed $T^2$ behavior and that the resistance in the
normal state did not depend on the external magnetic
field.\cite{jung,takano} Bud'ko \textit{et al.}~suggested that the clear
magnetoresistance might be much harder to detect in the samples
with defect scattering.\cite{budko1} Indeed, the residual
resistance ratio, RRR=$\rho$(300 K)/$\rho$(40 K), reported varies
with samples. The RRR reported by Ames Laboratory is about
25,\cite{budko1} much larger than those reported by other
groups,\cite{jung,takano} typically 2--3, even for epitaxial thin
films.\cite{kang1} It is worth noting that the superconducting
transition temperatures of the samples seem to be nearly the same
although their properties in the normal state are quite different.
To understand the nature of the resistivity and mangetoresistance
in the normal state, we report here a systematic study on
transport properties and magnetoresistance of bulk and thin-film 
samples with different nominal compositions. It is found that
these samples show almost the same transition temperature except
for the thin-film sample, while their temperature- and
magnetic-field-dependences of the normal state resistivity, as well as
their residual resistance ratios, are very different. We found that
the magnetoresistance is closely related to the residual
resistance ratio and can be scaled by a simple formula that can
be understood in terms of a reduction of the effective mean-free
path of the carrier in the presence of a magnetic field.

\section*{experiment}

The polycrystalline MgB$_{2}$ samples studied were prepared by a 
solid-state
reaction method. Small Mg chips (99.8\% pure) and boron powder
(99.7\% pure) with stoichiometries of Mg:B=1:2, 1.25:2, and 1.5:2 were
sealed inside Ta tubes in an Ar atmosphere. Each
sealed Ta ampoule was in turn enclosed in a quartz tube. The
ingredients were heated slowly up to 950 $^{\circ}$C and kept at this
temperature for 2 hr, followed by furnace-cooling to room
temperature. The samples so-prepared were granular and porous and
were used for measurements without further treatment. The thin-film
boron was prepared by sputtering. The thin-film MgB$_{2}$ was
synthesized by exposing thin-film boron to Mg vapor at 700~$^{\circ}$C
for adequate exposure time. Given that MgB$_{2}$ is the most Mg-rich
binary Mg-B compound known, it was felt that excess Mg would aid
in the formation of the proper, stoichiometric phase. It should be 
pointed out that most MgB$_{2}$ samples were synthesized with a 
stoichiometric starting
composition. However, noticeable amounts of MgO and/or Mg are either
frozen inside the samples or deposited on the walls of the sample 
containers.
Thus, the true Mg composition is certainly not the starting 
composition. Oxygen
contamination has also been widely discussed, particularly in terms 
of the
$T_c$ degradation of MgB$_{2}$ films.\cite{zhai} This could be why
the thin film has a very small RRR. The structure was
determined by powder X-ray diffraction, using a Rigaku DMAX-IIIB
diffractometer. The resistivity was measured using the standard
four-probe technique. The isothermal mangetoresistance was
measured using a SQUID magnetometer (Quantum Design). The magnetic
field was applied perpendicular to the measuring current direction. 
The transverse magnetoresistance in the normal state was measured.

\section*{Results and Discussion}

The powder X-ray diffraction (XRD) patterns of the samples show
the hexagonal MgB$_{2}$ phase, but with a minor amount of MgO as an
impurity. It is worth pointing out that no metal Mg is observed in
the XRD pattern for any sample, including those with excess Mg as
starting material. Figure \ref{fig1} shows the temperature dependence of
resistivity for all samples. For the bulk samples, the
superconducting transition temperature ($T_{c}$) is 38.3--39.3 K; $T_{c}$ of
the thin-film sample is 25 K. The residual resistance ratios (RRR)
of these samples are different, and the RRR is 3.03, 2.65, and 8.3
for the samples with the nominal composition Mg:B=1:2, 1.25:2, and
1.5:2, respectively; it is only 1.16 for the thin-film sample.  The
resistivity values at room temperature are 14, 88, and 49 $\mu\Omega$cm, 
respectively; and 188 $\mu\Omega$cm for the thin film. It has
been reported that the temperature dependence of resistivity
follows the $a+bT^3$ form reported for the polycrystalline MgB$_{2}$
with a RRR of 25,\cite{finnemore} while Jung \textit{et al.}~claimed 
that the
temperature dependence of resistivity follows $a+bT^2$ for the
high pressure sample with a RRR of 2.4.\cite{jung} To investigate
the resistivity behavior, the temperature dependence of
resistivity in the normal state for all samples has been fitted by
a formula $a+bT^c$, where $a$, $b$, and $c$ are fitting parameters. The
solid lines in Fig.~\ref{fig1} are the fitting curves by $a+bT^c$. The
experimental data could be fitted very well. These parameters are
listed in Table 1. The superconducting transition temperature
and width also show a systematic change, but their change is very
small, especially for $T_c$.

From Table 1, RRR increases with increasing
$x$. The fitting parameter $c$ also increases as $x$ increases, and 
is between 2.17 and 2.64, depending on the sample. These results indicate
that the RRR and the fitting parameter $c$ are closely related to
the nominal Mg content. The value of $c$
increases with RRR. This result is understandable from the empirical 
studies of conventional superconductors. Gurvitch has done
exhaustive studies on how the power law of $\rho$ {\it vs.} $T$ 
depends on
RRR or the residual resistivity.\cite{gurvitch} The power law
behavior in low temperature resistivity changes from $T^n$ ($n$=3--5)
for very clean samples to $T^2$ as disorder increases; the dirty
superconductors reach a limit $n$=2. Such a transition only takes
place in moderate- and strong-coupled superconductors. An example
of this effect is found in superconducting VN,\cite{zasadzinski}
which can be prepared with a wide variation in disorder, while
$T_c$ remains nearly constant and is quite insensitive to even
large amounts of disorder, similar to MgB$_{2}$. For the
well-ordered, stoichiometric VN film, the value of RRR is 8.4
($\rho_0=5.0$ $\mu \Omega cm$); while the RRR value drops to 1.14
($\rho_0$=63 $\mu \Omega cm$) for the same sample damaged by
radiation. The low temperature-dependent resistivity $\rho(T)$ from
10 to 30~K was fit to a function of the form $\rho_0+AT^n$. The
fitting parameter $n$ is 4 and 2.25 for the clean and dirty
(damaged) films, respectively. This is very similar to that
observed in MgB$_{2}$ and explains why the different
temperature-dependent resistivity is observed in MgB$_{2}$, as
discussed in the introduction. In order to compare with 
superconducting VN, the low temperature resistivity $\rho(T)$ from 
40 to 100~K is very well fit with the function of the form
$a_1+b_1T^{c_1}$ for all samples. It should be pointed out that
the fitting cannot be convergent for thin-film samples. Thus, the
$c_1$ is missing in Table 1. The results of the present study on
MgB$_{2}$ are consistent with the above, which might be another
indication that electron-phonon interaction is important. In
addition, the increased RRR and $c_1$ with excess Mg suggests
that the disorder was induced by Mg loss in the sample processing,
and excess Mg is needed to obtain more stoichiometric samples. In
fact, high quality MgB$_{2}$ wire was produced with much excess Mg
of a nominal ratio of Mg$_{2}$B.\cite{budko} The disorder could
arise from the Mg deficiency, which causes the scattering. In
addition, the high pressure study on MgB$_{2}$ has suggested
the existence of Mg non-stoichiometry.\cite{bordet} 

Figure \ref{fig2}a shows the isothermal transverse MR at 50~K for all
samples. It is found that the MR is always positive, and varies
with field in a similar fashion. It is interesting to note that
the value of magnetoresistance increases monotonously with
increasing residual resistance ratio. This accounts for the
huge MR observed by Ames Laboratory, while others observed almost zero 
MR; the Ames sample had a comparatively large RRR. It is
customary to analyze the classical orbital MR using the Kohler
plot. The motivation is that, in conventional metals, the
coefficient of the $B^2$ term is proportional to the transport
scattering time $\tau_{tr}(T)$. Since $\rho$ is proportional to
$1/\tau_{tr}(T)$, a slope of $\Delta \rho/\rho_0$ \textit{vs.}~$(H/\rho)^2$
should fall on a straight line with a slope that is independent of
$T$.\cite{pippard} The same data in Fig.~\ref{fig2}a are plotted on a Kohler
plot shown in Fig.~\ref{fig2}b. It is found that the $\Delta \rho/\rho_0$
\textit{vs.}~$H$ curve is a straight line at constant temperature for all
samples. The slope of $\alpha$ is between 1.6 and 2. It suggests that
the magnetic field dependence of mangetoresistance follows
$\Delta\rho(H)/\rho(0) \propto H^\alpha$ with $\alpha$=1.6--2,
which is nearly the same as $\alpha$ predicted by Kohler's rule.
Figure \ref{fig2}c shows a Kohler plot of magnetoresistance at 100~K and 
50~K for the sample with nominal composition Mg$_{1.5}$B$_2$. It is
found that the data fall on the single curve, independent of
temperature. These results suggest that there exists a single
salient scattering time in the normal state transport of MgB$_2$.
It should be pointed out that all samples with different nominal
composition follow the Kohler's rule although the RRR is
different. It implies that the normal state behavior of the
samples with different nominal composition arises from the crystal
grain rather than the grain boundary. Although the temperature
dependence of the resistivity for a ceramic sample often depends
on its porosity and inter-grain coupling, the large variation of
RRR suggests intrinsic differences in samples.

In order to clarify the differences in MR of samples with
different RRR,\cite{finnemore,jung,takano,budko1} we plot the
mangetoresistance at 50~K and under magnetic field of 5~Tesla \textit{vs.}~RRR, 
as shown in Fig.~\ref{fig3}. It is found that the MR increases
monotonously with increasing RRR. The data fit very well the
relation 0.04(RRR)$^{2.2\pm 0.1}$ as shown in Fig.~\ref{fig3}. It implies that
the MR could be scaled by the residual resistance ratio. From our
results, the mangetoresistance for the sample with RRR of 2--3
is only 0.2--0.5\%, while about 5\% for the sample with RRR of
8.3. For the sample with RRR of 25--26 a MR of about 50\% at
50~K and under 5~Tesla could be obtained from the formula above, consistent 
with that reported by Ames
Laboratory.\cite{finnemore,budko1} Our results provide an explanation
for the huge difference in magnetoresistance observed in MgB$_2$
by various groups. In fact, this behavior is understandable. It
has been pointed out that the magnetic field dependence of
isothermal magnetoresistance follows the Kohler's rule $\Delta
\rho/\rho_0$=const.$[H/\rho(0)]^2$. In a metal, the $\rho(0)
\propto 1/l$ and $l$ is the mean-free path, so that $\Delta
\rho/\rho_0 \propto (Hl)^2$. The RRR
is proportional to the mean free path $l$ at low temperature, so
that $\Delta\rho/\rho \propto$ RRR$^2$, which is in agreement
with our observation. It further suggests that the
mangetoresistance behavior arises from the crystal grain rather
than the grain boundary. The different results for the
magnetoresistance reported previously are thus attributed to the
different mean-free paths due to defect scattering in grains. A
significant sample-dependent defect scattering is expected. The
differences in the RRR, the temperature dependence of resistivity
in the normal state, and mangetoresistance for different samples have
been previously attributed to either the pressure\cite{kang} or
the thermal history\cite{cunningham} used during the synthesis.
However, the fact that similar low RRR and negligible
magnetoresistance have been observed in samples synthesized under
both ambient and high pressure challenges these
interpretations. The differences of resistivity and
magnetoresistance observed in this paper arise only from the
content of Mg in the samples, suggesting that the defect
scattering comes from the Mg deficiency in the sample. In fact,
Cooper \textit{et~al.}\cite{cooper} proposed in 1970 that boron in borides
AB$_2$ might be nonstoichiometric. In addition, Xue \textit{et
al.}\cite{xue} observed a correlation between the Mg loss and XRD
data as evidence for Mg deficiency.

\section*{conclusion}

The resistivity and mangetoresistance in the normal state have been 
investigated for samples with different RRR. The magnitude of 
mangetoresistance is found to vary with the residual resistance ratio, 
$a_0$(MR)=0.04(RRR)$^{2.2\pm 0.1}$. Although the transport properties of the 
samples are different, their superconducting transition temperatures are 
nearly the same. The mangetoresistance behavior follows the Kohler's rule. 
The different MR's and RRR's reported previously arise from defects in 
grain rather than the bound boundary, due to Mg deficiency. It is found
that the correlation between power law dependence of the low
temperature resistivity ($\rho$ {\it vs.} $T^c$) and the RRR is the same
as that observed in the strongly coupled superconductor VN.  Therefore,
it is understandable that the $\rho(T)$ behavior with different $c$
is observed in different groups, arising from the disorder (Mg
deficiency) in samples.

\section*{acknowledgments}

This work was supported in part by NSF Grant No. DMR-9804325, MRSEC/NSF Grant 
No. DMR-9632667, the T.~L.~L. Temple Foundation, the John and Rebecca Moores 
Endowment, and the State of Texas through the Texas Center for Superconductivity 
at University of Houston; and at Lawrence Berkeley Laboratory by the Director, 
Office of Energy Research, Office of Basic Sciences, Division of Material 
Science of the U.~S.~Department of Energy under Contract No.~DE-AC0376SF00098.

\begin{table}
\caption{Zero-resistance temperature ($T_c$), transition width
(TW), RRR of the samples, and the fitting parameter $c$ of
$a+bT^c$ to the temperature dependence of resistivity in the normal
state.}
\begin{tabular}{ c c c c c c}
sample   & $T_{c}(zero)$ (K)& TW  (K)& RRR &  $c$     &  $c_1$ \\ \hline
  MgB$_2$         & 38.3 & 0.8  & 3.03    & 2.17  &   2.87 \\
  Mg$_{1.25}$B$_2$  & 38.8 & 0.2  & 2.65    & 2.27  &   3.05 \\
 Mg$_{1.5}$B$_2$    & 39.3 & 0.08 & 8.30    & 2.43  &   3.41 \\
 thin film         & 25   & 2.3  & 1.16    & 2.64  &    -\\
\end{tabular}
\end{table}

\begin{figure}[tbp]
\caption{The temperature dependence of the normalized resistivity for
bulk and thin-film samples with nominal composition Mg$_x$B$_2$ 
($x$=1, 1.25, and 1.5). The solid lines are the fitting results of the 
formula $a+bT^c$ to experimental data.}
\label{fig1}
\end{figure}

\begin{figure}[tbp]
\caption{(a): The magnetic field dependence of isothermal magnetoresistance
at 50~K for samples with nominal composition Mg$_x$B$_2$ ($x$=1,
1.25, and 1.5) and thin film; (b): Kohler's plot for the same data
as Fig.~2a; the straight line is a guide line; (c): Kohler's
plot for isothermal magnetoresistance at 50 and 100~K for the
sample with nominal composition Mg$_{1.5}$B$_2$; the straight line
is a guide line.}
\label{fig2}
\end{figure}

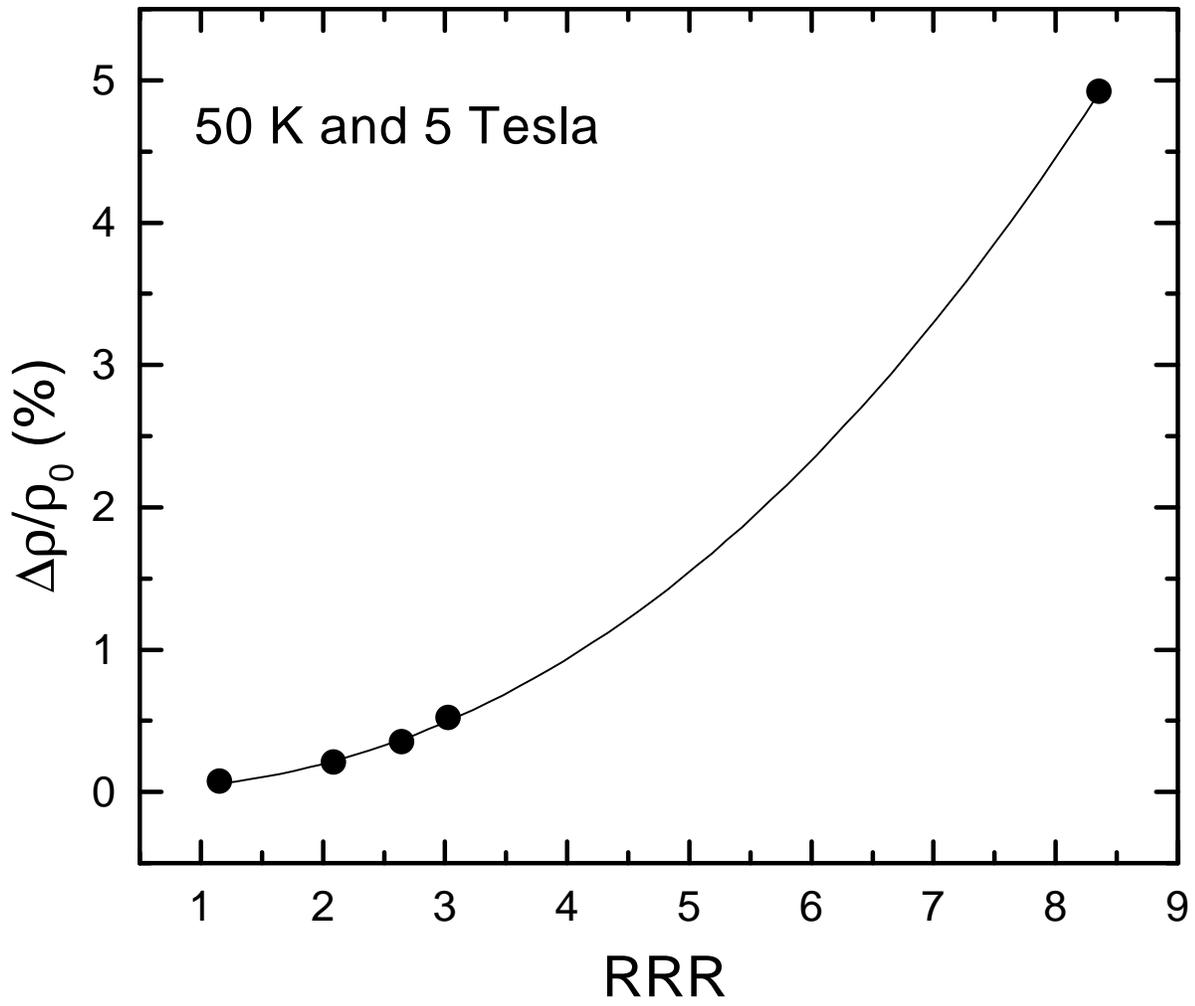
\begin{figure}[tbp]
\caption{The residual resistance ratio dependence of mangetoresistance at
50~K and under 5~Tesla; the solid line is a fitting curve of
0.04(RRR)$^{2.2\pm0.1}$ to experimental data.}
\label{fig3}
\end{figure}










\begin{references}

\bibitem{nagamatsu}
J. Nagamatsu, N. Nakagawa, T. Muranaka, Y. Zenitana and J. Akimitsu,
 Nature 410, 63 (2001).
\bibitem{kortus}
J. Kortus, I. I. Mazin, K. D. Belashchenko, V. P. Antropov and L. L. Boyer,
Phys. Rev. Lett. 86, 4656 (2001).
\bibitem{hirsch}
J. E. Hirsch, Phys. Lett. A 282, 392 (2001); J. E. Hirsch and F. 
Marsiglio, Phys. Rev. B 64, 144523 (2001).
\bibitem{budko}
S. L. Bud'ko, G. Lapertot, C. Petrovic, C. E. Cunningham, N. Anderson and 
P. C. Canfield,
Phys. Rev. Lett. 86, 1877 (2001).
\bibitem{lorenz}
B. Lorenz, R. L. Meng and C. W. Chu, Phys. Rev. B 64, 012507 (2001).
\bibitem{iwasa}
E. Saito, T. Takenobu, T. Ito, Y. Iwasa, K. Prassides and J. Arima, J. 
Phys.:
Condensed Matter 13, L267 (2001).
\bibitem{schmidt}
H. Schmidt, J. F. Zasadzinski, K. E. Gray and D. G. Hinks,
Phys. Rev. B 63, 220504 (2001).
\bibitem{sharoni}
A. Sharoni, I. Felner and O. Millo, Phys. Rev. B 63, 220508 (2001).
\bibitem{kremer}
R. K. Kremer, B. J. Gibson and K. Ahn,  cond-mat/0102432, February 23, 2001.
\bibitem{taka}
T. Takahashi, T. Sato, S. Souma, T. Muranaka and J. Akimitsu,
Phys. Rev. Lett. 86, 4915 (2001).
\bibitem{yidirim}
T. Yildirim, O. G\"{u}lseren, J. W. Lynn, C. M. Brown, T. J. Udovic,
Q. Huang, N. Rogado, K. A. Regan, M. A. Hayward, J. S. Slusky,
T. He, M. K. Haas, P. Khalifah, K. Inumaru and R. J. Cava,
Phys. Rev. Lett. 87, 037001 (2001).
\bibitem{kang}
W. N. Kang, C. U. Jung, K. H. Kim, M. Park, S. Y. Lee, H. Kim, E.
Choi, K. H. Kim, M. Kim and S. Lee,  Appl. Phys. Lett. 79, 982 (2001).
\bibitem{finnemore}
D. K. Finnemore, J. E. Ostenson, S. L. Bud'ko, G. Lapertot and P. C.
Canfield, Phys. Rev. Lett. 86, 2420 (2001).
\bibitem{jung}
C. U. Jung, M. S. Park, W. N. Kang, M. S. Kim, K. H. Kim, S. Y. Lee
and S. I. Lee, Appl. Phys. Lett. 78, 4157 (2001).
\bibitem{takano}
Y. Takano, H. Takeya, H. Fujii, H. Kumakura, T. Hatano, K. Togano, H. Kito 
and 
H. Ihara,
Appl. Phys. Lett. 78, 2914 (2001).
\bibitem{budko1}
S. L. Bud'ko,  C. Petrovic, G. Lapertot, C. E. Cunningham, P. C. 
Canfield, M. H. Jung and A. H. Lacerda,
Phys. Rev. B 63, 220503 (2001).
\bibitem{kang1}
W. N. Kang, H. J. Kim, E. M. Choi, C. U. Jung and S. I. Lee,
Science 292, 1521 (2001).
\bibitem{zhai}
H. Y. Zhai, H. M. Christen, L. Zhang, M. Paranthaman, C. Cantoni, 
B. C. Sales,
P. H. Fleming, D. K. Christen and D. H. Lowndes, cond-mat/0103618, 
March 30, 2001.
\bibitem{gurvitch}
M. Gurvitch, Phys. Rev. Lett. 56, 647 (1986).
\bibitem{zasadzinski}
J. F. Zasadzinski, A. Saggese, K. E. Gray, R. T. Kampwirth and R. Vaglio,
Phys. Rev. B 38, 5065 (1988).
\bibitem{bordet}
P. Bordet, M. Mezouar, M. Nu\~{n}ez-Regueiro, M. Monteverde, 
M. D. Nu\~{n}ez-Regueiro, N. Rogado, K. A. Regan, M. A. Hayward, T. He, 
S. M. Loureiro and R. J. Cava, cond-mat/0106585, June 28, 2001.
\bibitem{pippard}
A. B. Pippard, Magnetoresistance in Metals (Cambridge UK: Cambridge, 1989).
\bibitem{cunningham}
C. E. Cunningham, C. Petrovic, G. Lapertot, S. L. Bud'ko, F. Laabs, 
W. Straszheim, D. K. Finnemore and P. C. Canfield, Physica C 353, 5 
(2001).
\bibitem{cooper}
A. S. Cooper, E. Corezwit, L. D. Longinotti, B. T. Matthias and W. H. 
Zachariasen, Proc. Natl. Acad. Sci. 67, 313 (1970).
\bibitem{xue}
Y. Y. Xue, R. L. Meng, B. Lorenz, J. K. Meen, Y. Y. Sun and C. W. Chu, 
cond-mat/0105478, May 24, 2001.
\end{references}
\end{document}